\pdfoutput=1

\documentclass[11pt]{article}

\usepackage[final]{ACL2023}

\usepackage{times}
\usepackage{latexsym}
\usepackage{graphicx}

\usepackage[T1]{fontenc}

\usepackage[utf8]{inputenc}

\usepackage{microtype}
\usepackage{booktabs}
\usepackage{tabularx}
\usepackage{amsmath}

\usepackage{inconsolata}
\usepackage{listings}
\usepackage[dvipsnames]{xcolor}

\title{Is Passive Expertise-Based Personalization Enough?\\A Case Study in AI-Assisted Test-Taking}

\author{
 \textbf{Li Siyan\textsuperscript{1}\footnotemark[1]},
 \textbf{Jason Zhang\textsuperscript{2}\footnotemark[1]},
 \textbf{Akash Maharaj\textsuperscript{3}},
 \textbf{Yuanming Shi\textsuperscript{3}},
 \textbf{Yunyao Li\textsuperscript{3}}
\\
\textsuperscript{1} Columbia University, \textsuperscript{2} Georgia Institute of Technology, \textsuperscript{3} Adobe
\\ \footnotemark[1] ~~Work done during internship at Adobe
}

\begin{document}
\maketitle
\begin{abstract}
Novice and expert users have different systematic preferences in task-oriented dialogues. However, whether catering to these preferences actually improves user experience and task performance remains understudied. To investigate the effects of expertise-based personalization, we first built a version of an enterprise AI assistant with passive personalization. We then conducted a user study where participants completed timed exams, aided by the two versions of the AI assistant. Preliminary results indicate that passive personalization helps reduce task load and improve assistant perception, but reveal task-specific limitations that can be addressed through providing more user agency. These findings underscore the importance of combining active and passive personalization to optimize user experience and effectiveness in enterprise task-oriented environments.



\end{abstract}

\section{Introduction}

Personalization has been heralded as a silver bullet for improving conversational agents \cite{kocaballi2019personalization, ait2023power}. Tailoring responses to user preferences have been shown to improve user engagement \cite{ha2024clochat}, foster positive attitude \cite{rhee2020effects}, and enhance task efficiency \cite{joshi2017personalization}. In terms of relevant user information, prior personalized systems have leveraged dimensions such as personality \cite{ait2023power}, conversation history \cite{li2024hello}, demographics \cite{zheng2019personalized,joshi2017personalization}, medical information \cite{abbasian2023conversational}, writing style \cite{salemi2023lamp,li2024learning}, and task-specific preferences \cite{joko2024doing,luo2019learning}. 

\begin{figure}
    \centering
    \includegraphics[width=0.8\linewidth]{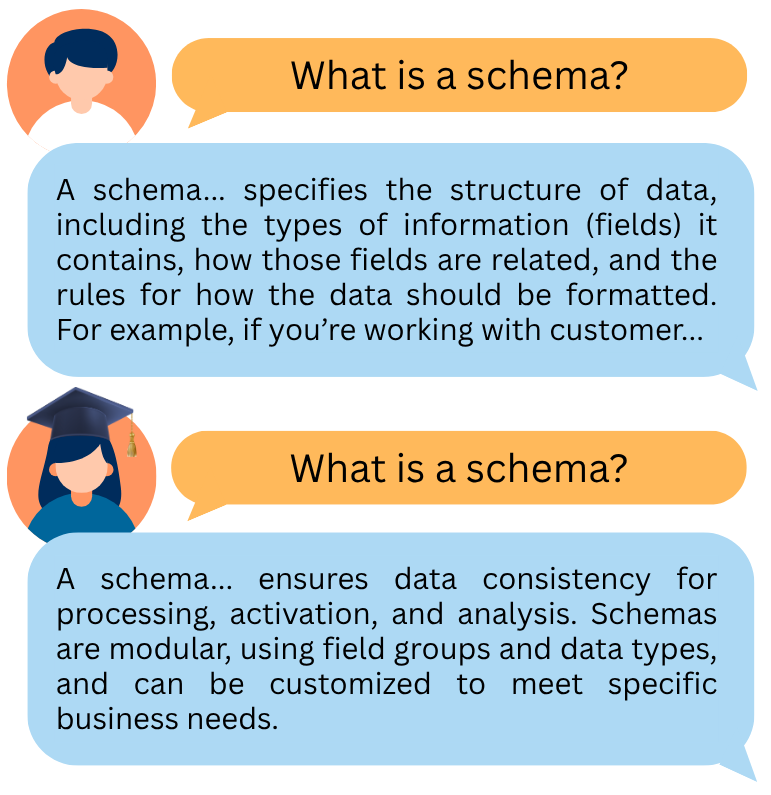}
    \caption{Example responses to the query ``What is a schema'' for a novice (top) and an expert (bottom).}
    \label{fig:utt_examples}
\end{figure}

One crucial personalization factor often under-discussed is \textbf{user expertise} in task-oriented settings.
Users with different expertise levels require different accommodations from task-oriented conversational assistants \cite{jokinen-kanto-2004-user}. Novice users may prefer a response that is easier to understand and contains more step-by-step instructions, whereas expert users may opt for concise explanations with more jargon \cite{janarthanam-lemon-2014-adaptive}. 
Figure \ref{fig:utt_examples} illustrates example responses for a novice and an expert: the response for the novice is more comprehendible and augmented with an example, while that for the expert is succinct and jargon-dense.

Personalizing on user expertise may not only address user preferences, but may additionally narrow the novice-expert gap in human-AI collaborative tasks. Experts can often benefit more from AI explanations as experts are more capable of disregarding hallucinations \cite{inkpen2023advancing,chen2024learning}. Early work suggests that adapting responses to user expertise can reduce novice users' acclimation time \cite{komatani2003flexible}, improve task efficiency \cite{janarthanam-lemon-2014-adaptive}, and increase user engagement \cite{palta2025speaking}. However, how expertise-based personalization interplays with the success of \textit{knowledge-intensive tasks} has not been systematically studied. 


With the recent advancements in Large Language Models (LLMs), personalization can be easily incorporated through prompt adjustments \cite{zhang2024personalization}. We see this as a unique opportunity to study the effects of expertise-based personalization in knowledge-intensive tasks through designing interpretable personalization guidelines through principled prompting.

As an initial exploration, we target the task of question-answering from product documentation in an enterprise AI assistant on a customer data management platform, which we will refer to as Platform A for anonymity. We augment this AI assistant with the capability of passive expertise-based personalization, leveraging detected and pre-defined user expertise. Due to the diversity of users on this platform, relevant product knowledge spans an extensive set of domains, and one user can easily be an expert in one domain but a novice in another. As a result, in addition to utilizing the overall user knowledge, we utilize domain-level expertise in personalization, consistent with previous systems \cite{jokinen-kanto-2004-user,jokinen2004evaluation}.

We conducted a pilot user study to investigate whether catering to expertise-specific user preferences improves user task performance and conversational experience. Participants completed two sets of \textbf{timed} exam questions under AI assistance, with the assistant being personalized to their self-reported expertise for only one of the question sets.

We find that expertise-based personalization \textit{improves aspects of task load}, such as subjective task success and temporal demand, and \textit{enhances assistant helpfulness and relevance}. However, under time pressure, non-experts do not always benefit from the personalized, more informative responses, even if they would prefer them in less-pressured contexts. This accentuates the value of active personalization, allowing users to adjust assistant responses in a task-dependent manner.


Our contributions include the following:

\begin{enumerate}
    \item We design an extensible framework for passive expertise-based personalization.
    \item We examine the effect of such personalization using a small-scale, within-subject user study, where we identify task-specific limitations of our passive personalization approach.
    \item We explore qualitative and quantitative differences in expert and non-expert behaviors.
\end{enumerate}

\section{Related Work}
Prior expertise-based personalization work primarily adopts passive personalization, where users are granted little steering power besides self-reporting expertise. 
\citet{jokinen-kanto-2004-user} adapts a speech-based email system, AthosMail, using timeouts, the number of help requests, and interruptions. The follow-up work \cite{jokinen2004evaluation} suggests that the types of user errors can be indicative of expertise level. \citet{maloor2000dynamic} develops a help-desk system that maintains local and accumulated expertise levels through system metrics including response complexity, goal complexity, and accomplishment time. \citet{komatani2003flexible} uses acoustic cues to categorize user skill level and knowledge levels of an interactive bus information system. \citet{wilensky1987berkeley} assesses user expertise by assuming a knowledge-expertise mapping, illustrating responses with examples when the user is a novice. In all of these systems, responses to novices tend to have more detail, while those for experts are more technical. One notably different adaptation approach is a reinforcement learning-based method in \citet{janarthanam-lemon-2014-adaptive}, where the authors study a human-assistant collaborative assembly task. The assistant's use of technical jargon is kept consistent with the user's. Results indicate that adapting responses to user technical jargon knowledge improves user task success. There remains limited literature examining the \textit{effects} of expertise-based personalization in knowledge-intensive tasks, highlighting the need to revisit such approaches in the era of LLMs.

Recent work has explored implicit expertise classification using conversational data, including an LLM-driven framework trained using self-reported expertise and assessment responses \cite{David2025ProfiLLMAL}, 
a vocabulary-centered deep-learning model \cite{ferrod2021identifying}, and prompting strategies to classify perceived expertise of both the user and the AI assistants \cite{palta2025speaking}. Given the scale of our user study and limited ground-truth data, we prioritize data-efficient methods and leave advanced classification approaches for future work. Additional approaches for inferring user expertise include student knowledge tracing \cite{neshaei2024towards, scarlatos2025exploring, jung2024clst}, as well as through search behavior \cite{zhang2015predicting, kiseleva2015impact}. We do not discuss these approaches as they are not inherently conversational. 






\begin{figure*}[!ht]
    \centering
    \includegraphics[width=\textwidth]{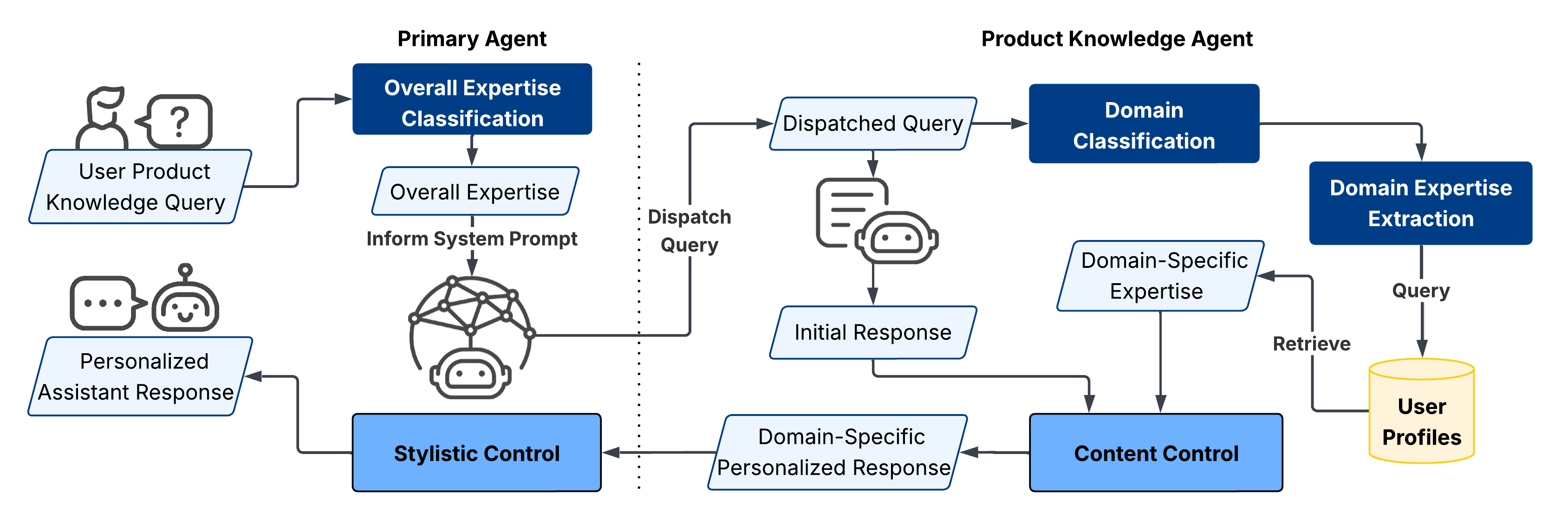}
    \caption{The overview of our system, capable of fine-grained passive personalization.}
    \label{fig:overview}
\end{figure*}

\section{Expertise-based Personalization Implementation}

The enterprise AI assistant we leverage for this work follows an agentic design. 
Our specific focus is to perform expertise-based personalization for both (1) a product knowledge agent and (2) the primary agent.

Figure \ref{fig:overview} describes the overall flow of our system. When a user submits a product-knowledge query, the primary agent starts by estimating the user's overall expertise level (Section \ref{sec:expertise_classifier_desc}) and then dispatches a slightly reworded query to the product-knowledge (PK) agent. Within the PK agent's workflow, a domain classifier determines the query domain (Section \ref{sec:domain_classifier_desc}), and the user's domain-level expertise is retrieved. Using this information, the PK agent both personalizes and includes the \textit{domain-specific} expertise level in its response. The primary agent would then incorporate stylistic constraints using \textit{overall} expertise (Section \ref{sec:response_adaptation_desc}) to craft the final response.

In the following sections, we mainly utilize a collection of 1,785 internal user queries for the AI assistant, collected during 2024 - 2025. For LLM usage, unless otherwise specified, we used \texttt{GPT-4o}.

\subsection{Domain Classification}
\label{sec:domain_classifier_desc}

Currently, the product knowledge agent is implemented using Retrieval-Augmented Generation \cite{lewis2020retrieval}. A noisy mapping between the query and its domain(s) can be achieved through the domains of the retrieved documents. 

We constructed an LLM-based classifier for query domains. The input of the classifier includes the query and high-level domain descriptions, and the output consists of a single prediction. 
To obtain a good classification prompt without ad hoc engineering, we experimented with the SIMBA optimizer\footnote{\url{https://dspy.ai/api/optimizers/SIMBA/}} from the DSPy prompt optimization framework \cite{opsahl-ong-etal-2024-optimizing}. The optimization process aims to maximize classification accuracy (i.e. is the predicted domain among one of the silver-truth document domains). We randomly selected 300 queries as the training set and used the rest as the validation set, according to DSPy official recommendations. Using this noisy data, we achieved validation top-1 accuracy of $82.2\%$. 

\subsection{Expertise Classification Dimensions}
\label{sec:expertise_classifier_desc}

Prior work has often relied on self-reported expertise levels to adapt interactive system responses. We are interested in exploring on-the-fly options for accurate expertise classification to remove the burden of self-reporting, focusing on the following classification dimensions:

\noindent \textbf{LLM Judgment:} \citet{palta2025speaking} created a prompt to classify user utterances in Bing Chat conversations into five tiers of user expertise. For our purposes, we reduce these tiers to only include Novice (0), Intermediate (1), and Expert (2), and modified the prompt accordingly (Appendix \ref{app:expertise_prompt}).

\noindent \textbf{Words-per-sentence Ratio:} \citet{kim2011experts} finds that words-per-sentence ratios are significantly different between cultural event evaluations written by domain experts and novices. This generally aligns with our intuition that experts are capable of more nuanced reasoning of domain knowledge, which requires more verbiage to articulate. 

\noindent \textbf{Jargon Usage:} Jargon expressions can be indicative of user technical expertise \cite{janarthanam-lemon-2014-adaptive,ferrod2021identifying}. To estimate this, we reference an official dictionary for Platform A terminologies, and compute the number of terms present for each utterance through fuzzy string matching\footnote{\url{https://github.com/seatgeek/thefuzz}}.

\noindent \textbf{Readability:} Prior work, such as \citet{toma2015tell}, notes that the length of words affects the perceived expertise of messages, with longer words being associated with more expertise. The length of words (can be approximated as the number of syllables per word) and the length of utterances together inform readability \cite{flesch2007flesch,dale1948formula}. We use an existing readability calculator library to obtain a set of nine readability scores\footnote{\url{https://pypi.org/project/readability/}}. 


We analyzed a dataset containing 1,198 internal, use-case-specific queries, each labeled as originating from intermediate (781 queries) or expert users (417 queries). To assess the discriminative power of these dimensions, we performed independent t-tests to contrast the two groups (Table \ref{tab:expertise_diff}). The results reveal statistically significant differences along multiple dimensions, suggesting that our classification scheme effectively distinguishes between intermediate and expert queries. 
As expected, expert users produce longer utterances and more jargon, and are classified by the LLM judge to exhibit more expertise. Notably, their queries are also more readable. Extrapolating from this, we assume that novice queries would have lower word-per-sentence ratios, less jargon usage, and novice-level LLM-judged expertise.

\begin{table}[!h]
    \centering
    \begin{tabularx}{\linewidth}{c|cc}
    \toprule
    \textbf{Dimension} & \textbf{Intermediate} & \textbf{Expert} \\
    \midrule
    LLM Judgment & 0.431* & 0.528* \\
    Word Ratio & 9.66** & 11.0** \\
    Jargon & 0.104** & 0.194** \\
    Readability & 79.3** & 86.0** \\
    \bottomrule
    \end{tabularx}
    \caption{Mean values along each classification dimension of our intermediate and advanced queries. For readability, we select the Flesch Reading Ease score (*$p<0.05$, **$p<0.001$).}
    \label{tab:expertise_diff}
\end{table}

These dimensions will be employed to analyze queries from our user study. If similar trends persist, we can more reliably utilize these dimensions for passive expertise-based personalization. 



\subsection{Response Adaptation}
\label{sec:response_adaptation_desc}

Balancing explainability against resource and implementation constraints, we choose prompt modification over fine‑tuning to shape assistant responses.


\subsubsection{Content Control}
We reference an established framework for different instruction strategies as students gain domain expertise \cite{persky2017moving}, as well as observations from \citet{wilensky1987berkeley} and \citet{janarthanam-lemon-2014-adaptive}. See more details of the framework in Appendix \ref{app:instruction_strats}.

A paraphrase of the following directive is roughly incorporated into both the primary agent's system prompt and a new developer prompt used for PK agent post-processing:

\begin{lstlisting}
1. If the user has Novice expertise level:
- Provide basic and straightforward explanations, avoid using advanced jargon
- Be more detailed in your explanations
- Give step-by-step directions
- Help the user organize knowledge through tables, if and when appropriate
2. If the user has Expert expertise level:
- Provide more domain-specific, deeper knowledge
- You can use more jargon when appropriate
- Be succinct and direct in your descriptions
3. If the user has Intermediate expertise level:
- Balance your response between the Novice and the Expert specifications
\end{lstlisting}

\subsubsection{Stylistic Control}
During initial testing, we noticed that the personalized AI assistant can still not comply with specified stylistic preferences in the prompts. Consequently, we devise stylistic control developer prompts for the primary agent. We focus on the following dimensions: \textit{Readability}, as novices should receive more readable responses \cite{joshi2025eli}; \textit{Jargon density}; and \textit{Message Length}. The developer prompt is injected in the following manner:

\begin{lstlisting}
[Original System Prompt]
---
[Developer Prompt]
---
User Preferences: [User Preferences]
\end{lstlisting}

We assume that novices prefer responses to be \textit{more readable and have less jargon}, while experts would have the opposite preferences. We additionally assume that experts prefer \textit{shorter} responses \cite{jokinen-kanto-2004-user}. 

To create a sufficiently robust developer prompt, we leverage DSPy's MIPROv2 optimizer and \texttt{GPT-4o-mini} to propose $15$ prompts and evaluate them on $150$ internal queries. We additionally sample from our assumed expert and novice preferences to pair with the queries. 

We use \textbf{compliance} of preferences as our metric. Specifically, for a given user query, we would generate two responses, the original personalized response $R_{OG}$ and the response with the developer prompt $R_{D}$. Then, we contrast their values for the stylistic dimensions. For example, if the preference is \textit{more readable}, and $\texttt{Readability}(R_{D}) > \texttt{Readability}(R_{OG})$, then we mark this response as compliant. If the preference is \textit{no preference}, compliance is always satisfied.

The best prompt is a Pareto-optimal prompt with the highest averaged percentage compliance, as this is a multi-objective optimization problem. We document the compliance scores in Table \ref{tab:prompt_opt}.

\begin{table}[!h]
    \centering
    \begin{tabular}{c|ccc}
    \toprule
    \textbf{Prompt} & \textbf{Read.} & \textbf{Jargon} & \textbf{Length}\\
    \midrule
        Original & 53.1 & \textbf{78.1} & 100.0\\
        Best Proposed & \textbf{75.8} & 75.8 & \textbf{100.0}\\
    \bottomrule
    \end{tabular}
    \caption{Percentage compliance scores for the two developer prompts along the stylistic control dimensions.}
    \label{tab:prompt_opt}
\end{table}

We see that compliance for \textit{Readability} and \textit{Jargon Density} is lower than for \textit{Length}, likely because increasing jargon and reducing readability is challenging in shorter responses, as Flesch reading ease depends on both word and sentence length.

\section{User Study}


We recruit 16 users of Platform A through internal channels. Participants indicated their roles and experience with the platform in a pre-survey. To enable finer-grained adaptation without excessively questioning our participants, we develop a role-expertise mapping by consulting experts to help us infer the domain-level expertise level of an average person with a specific role. We assume that the expertise level is known and static throughout the interaction to reduce cascading errors from our expertise classifiers. Our study aims to answer these research questions:

\noindent \textbf{RQ1:} Does passive expertise-based personalization improve task performance?

\noindent \textbf{RQ2:} Does passive expertise-based personalization improve conversational experience?

\noindent \textbf{RQ3:} Do the expertise classification dimensions remain successful in distinguishing between expert and novice user queries in our task?

\subsection{Material Preparation}
For our specific task, participants are asked to complete two sets of eight product knowledge questions about Platform A. These questions are sampled from 24 manually-curated questions from publicly available certification exams.

To ensure the evenness of question difficulty among the two sets of questions, we recruited two expert Platform A users to rate their difficulty. These experts resolved disagreements through discussion. After removing questions where significant disagreements remain, the resulting difficulty ratings exhibit a Krippendorff's $\alpha$ value of $0.899$. See more details of the selection and annotation process in Appendix \ref{app:material_preparation}. Stratified sampling is then performed to ensure that both sets contain the same number of questions for each difficulty level.

\subsection{Experimental Procedure}

A version of the AI assistant and a test interface are presented to the participant side by side, with the window containing the AI assistant placed on the left. To avoid order effects, a researcher flips a coin to decide if the personalized or baseline assistant is presented first. Participants were allowed 10 minutes to complete each set of certification questions. For each question, the participant must answer a quick follow-up (Table \ref{tab:follow_up}). These additional instructions were provided:

\begin{quote}
 \textit{Ask the AI assistant about anything you are not certain about. Avoid copying the questions directly into the AI assistant.}
\end{quote}

\begin{table}[]
    \centering
    \begin{tabularx}{\linewidth}{X}
    \toprule
    \textbf{Q:} How did you answer this question?\\
    \midrule
     A. I knew the answer\\
     B. I guessed the answer\\
     C. The AI Assistant gave me the answer directly\\
     D. I was able to extrapolate the answer from assistant response\\
     \bottomrule
    \end{tabularx}
    \caption{Follow-up question after each exam question.}
    \label{tab:follow_up}
\end{table}

Note that while the two sets of problems are fixed, the individual question sets are shuffled. Upon completion of every set of eight questions, the participants completed a brief post-survey to indicate their experience. In addition to the \textbf{NASA Task Load Index (NASA-TLX)} \cite{hart1988development}, the post-survey contains 5-point Likert scale questions about \textbf{response quality}:
\begin{enumerate}
    \item How helpful are the assistant responses? (1 - Not helpful at all; 5 - Very helpful)
    \item Are you able to understand all of the assistant responses? (1 - I had some difficulty understanding all of the assistant responses; 5 - I had no difficulty)
    \item How relevant are the assistant responses? (1 - Not relevant; 5 - Very relevant)
    \item How well do you think the assistant's responses reflect your expertise level on the areas you asked about? (1 - Severely misaligned; 5 - Very well-aligned)
\end{enumerate}

\section{Results and Discussion}

Three experts completed the exams unaided, and we remove their data from our analyses. This results in a total of 13 participants in our final analyses, comprising of \textbf{five novices, four intermediates, and four experts}. For the specific user roles, one person is in administration, one in product, and the rest are in engineering.

\begin{figure*}[!ht]
    \centering
    \includegraphics[scale=0.4]{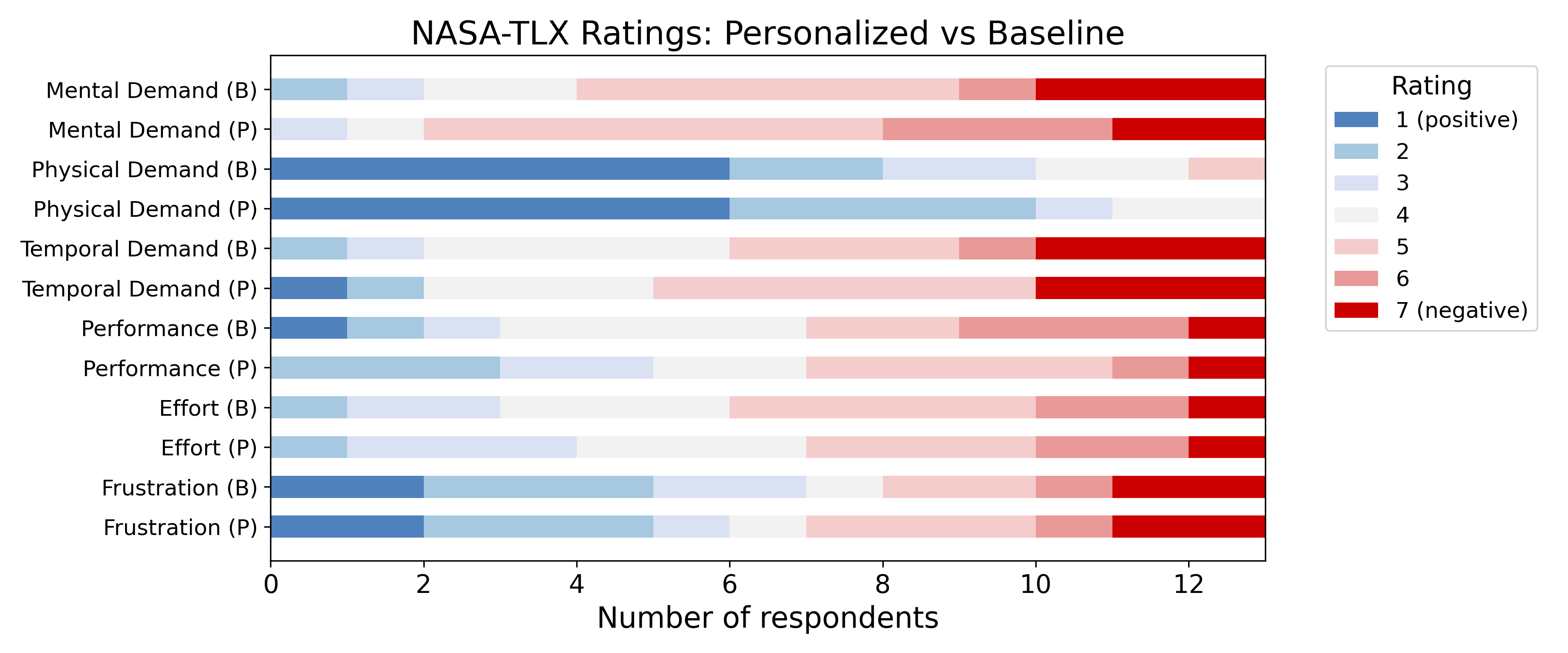}
    \begin{tabular}{c|cccccc}
    \toprule
         \textbf{Condition} & \textbf{Mental} $\downarrow$ & \textbf{Physical} $\downarrow$ & \textbf{Temporal} $\downarrow$ & \textbf{Performance} $\downarrow$& \textbf{Effort} $\downarrow$& \textbf{Frustration} $\downarrow$\\
         \midrule
         Baseline & \textbf{5.00} & 2.23 & 4.85 & 4.38 & 4.53 & \textbf{3.69}\\
         Personalized & 5.31 & \textbf{1.92} & \textbf{4.69} & \textbf{4.08}& \textbf{4.38}& 3.84\\
         \bottomrule
    \end{tabular}
    \caption{NASA-TLX results for baseline (B) and our personalized system (P) for all 13 included participants. We report the average ratings in the table.}
    \label{fig:nasa_tlx}
\end{figure*}

\subsection{Exam Performance}

\noindent \textbf{Exam Scores:} We are interested in \textit{how personalization affects exam scores}. As previously noted, the exam question sets are matched in difficulty according to expert annotations. Furthermore, since the personalized assistant is built upon the baseline version, the core response content should be similar. We thus can attribute exam performance difference to the effects of personalization with reasonable confidence. In our analysis, we target questions where participants consulted the AI assistant. The percentage score for each participant is computed:
\begin{center}
    $\%\text{Score} = \frac{\#(\text{Correct Answers})}{\#(\text{Questions with AI Assistance})}$
\end{center}

Table \ref{tab:test_scores} documents the average exam percentage scores across expertise levels and personalization conditions. Personalization may improve scores for novices and experts, but not intermediate users. Our preliminary finding is consistent with \citet{inkpen2023advancing}: intermediate users have high variance when receiving AI recommendations, while novices and experts improve more consistently.

\begin{table}[!h]
\centering
\begin{tabular}{c|ccc|c}
\toprule
    \textbf{Cond.} & \textbf{Novice} & \textbf{Inter.} & \textbf{Expert} & \textbf{Overall}\\
    \midrule
    B & 49.2 & \textbf{56.3} & 68.3 & 58.1\\
    P & \textbf{56.0} & 51.3 & \textbf{74.2} & \textbf{62.0}\\
    \bottomrule
\end{tabular}
\caption{The average exam percentage scores for questions where the AI assistant is used.}
\label{tab:test_scores}
\end{table}

\noindent \textbf{Assistant Response Usage:} Participants sent the baseline assistant a total of 88 inquiries, and personalized assistant 89 inquiries. We intend to examine \textit{how} participants used the AI assistant responses. During our task, participants indicated how they leveraged the responses through follow-up questions. We compute the percentages for each of the options except ``I knew the answer'' (Table \ref{tab:followup_percents}). 

\begin{table}[!h]
    \centering
    \begin{tabular}{c|ccc}
    \toprule
        \textbf{Cond.} & \textbf{\%Guess} & \textbf{\%Direct} & \textbf{\%Extrapolate} \\
        \midrule
        B & 34.5 & 47.3 & 18.2 \\
        P & 42.3 & 23.1 & 34.6 \\
         \bottomrule
    \end{tabular}
    \caption{Proportions per condition (baseline and personalized) for each of the follow-up response options: Guessed the answer (\%Guess), directly given the answer (\%Direct), and extrapolated the answer (\%Extrapolate). }
    \label{tab:followup_percents}
\end{table}

We observe that participants are able to extrapolate responses from AI assistant responses more with passive personalization. Additionally, we find that non-expert participants guess slightly less with personalization (28.2\% versus 29.5\%). This \textit{potentially} indicates that expertise-based personalization can foster deeper thinking, as participants can comprehend the personalized responses sufficiently to extrapolate. However, it is not clear whether encouraging more mental processing of assistant responses would always be desirable, since users may sometimes prefer more straightforward answers.

\subsection{Task Load}

Figure \ref{fig:nasa_tlx} documents the NASA-TLX metrics results. We note that, in general, the personalized assistant \textit{reduces} physical and temporal demand, \textit{improves} subjective task success assessment, and \textit{decreases} perceived effort. However, participants indicate higher mental load with the personalized assistant, as well as higher frustration. We include NASA-TLX results for non-experts in Appendix \ref{app:task_performance_ne}, which indicate similar patterns, with even more significant increase in mental demand (a gap of $0.44$ compared to $0.31$).

A plausible explanation for the personalized assistant's worse performance in alleviating mental load and frustration is its response verbosity and latency for novices, which can be more overwhelming in our time-constrained setting. The latency stems from the additional LLM calls and longer prompts in our system design, and novice-oriented responses are more verbose as the assistant aims to incorporate more examples and explanations.

\subsection{Conversational Experience}

We record the average ratings of assistant response quality in Table \ref{tab:assistant_response}. Although the personalized assistant is rated as more helpful and relevant overall, participants report greater difficulty in understanding all of its responses and perceive the personalized responses as less accurately aligned with their own expertise levels. The lower understandability scores may again be explained by the verbosity of the personalized assistant.

\begin{table}[!h]
\centering
\begin{tabular}{c|cccc}
\toprule
    \textbf{Cond.} & \textbf{Help.} & \textbf{Understand.} & \textbf{Rel.} & \textbf{Expert.}\\
    \midrule
    B & 2.84 & \textbf{3.61} & 3.00 & \textbf{3.00}\\
    P & \textbf{3.08} & 3.46 & \textbf{3.07} & 2.92\\
    \bottomrule
\end{tabular}
\caption{The average response quality ratings for the baseline (B) and personalized (P) assistants: Helpfulness, Understandability, Relevance, and Expertise Alignment Accuracy.}
\label{tab:assistant_response}
\end{table}

For expertise alignment accuracy, expert users in fact consider the personalized assistant better aligned (Table \ref{tab:assistant_response_exp}), while non-experts prefer the baseline. We hypothesize that, despite reducing jargon, the personalized responses for non-experts may still contain advanced terms. Without an official taxonomy of jargon difficulty for Platform A, we currently lack the means to further simplify these personalized responses.

We attempt to address the confounding factor of response verbosity in a time-constrained task setting through our follow-up study (Section \ref{sec:follow_up_novice}).

\subsection{Follow-up Novice Preference Study}
\label{sec:follow_up_novice}


To investigate novice preferences under more stress-free conditions, we asked four of our novice participants to indicate pairwise preferences when comparing the baseline and the personalized AI assistant responses for the same queries. For each novice user, we randomly sampled three to four of their context-independent queries during the user study where the AI assistant was helpful, and asked them to select the better response along the same assistant response quality axes. We selected a small query sample set to keep the annotation task manageable, and because it is difficult for either version of the AI assistant to answer certain novice queries due to the ambiguous nature of these queries. We include the preference rates of the responses in this setting in Table \ref{tab:novice_prefs}.

\begin{table*}[!h]
    \centering
    \begin{tabular}{c|c|cc|cccc}
    \toprule
         \textbf{Expertise} & \textbf{\# Queries} & \textbf{Partial Sim.} & \textbf{Sem. Sim.} & \textbf{LLM Judge} & \textbf{Length} & \textbf{Jargon} & \textbf{Read.} \\
         \midrule
         Novice & 17.6 & 74.7* & 0.720 & 0.670* & 20.1** & 1.97* & 63.8 \\
         Intermediate & 12.0 & 72.8* & 0.674 & 0.312* & 9.31** & 1.60*& 55.7\\
         Expert & 10.3 & 67.5* & 0.673 & 0.439* & 10.8** & 1.22*& 70.5\\
         \bottomrule
    \end{tabular}
    \caption{The average metric values for queries corresponding to different expertise levels. Statistical significance established through one-way ANOVA (*$p<0.05$, **$p<0.001$).}
    \label{tab:diff_exp}
\end{table*}

\begin{table}[!h]
    \centering
    \begin{tabular}{c|cccc}
    \toprule
        \textbf{Cond.} & \textbf{Help.} & \textbf{Understand.} & \textbf{Rel.} & \textbf{Expert.} \\
        \midrule
         B & 21.4 & 14.3 & 21.4 & 14.3\\
         P & \textbf{78.6} & \textbf{85.7} & \textbf{78.6} & \textbf{85.7}\\
         \bottomrule
    \end{tabular}
    \caption{Preference rates of the personalized assistant (P) against the baseline assistant (B) in our follow-up novice preference study.}
    \label{tab:novice_prefs}
\end{table}

The results indicate that \textbf{novices significantly prefer the personalized responses under a more relaxed setting}. 
Novice participants mentioned that the personalized responses provide helpful examples and step-by-step directions, and highlight the ``why'' about selecting the correct multiple-choice options. They additionally enjoy that the personalized assistant provides more context and explanations in an easily digestible manner, catered to their expertise.

Our results here suggest an interaction between the nature of the task and user preference outcomes. When the novice users are not as time-constrained, they are able to read through the lengthier personalized responses and reap the benefits; in contrast, when they are pressed for time, novices may gravitate toward more concise responses, even when those are not as informative. These findings emphasize the importance of considering task-related factors when designing personalized experiences, and a need to offer users active control over stylistic aspects of assistant responses, such as toggling for response lengths.

\subsection{Differences between Novices and Experts}
We identify quantitative and qualitative differences in how experts and non-experts interact with the assistant, and document results in Table \ref{tab:diff_exp}. 

\noindent \textbf{Qualitative Differences:} In terms of purposes of using the assistant, \textit{novices and intermediates} use it mainly as an \textit{educational resource} when unsure about answers or questions, while \textit{experts} primarily use it as a \textit{sanity check}. Experts are more likely to recognize and ignore AI assistant hallucinations, whereas non-expert users often do not. These observations align with existing literature. 

\noindent \textbf{Quantitative Differences:} To begin with, \textit{novices asked more questions}, likely since novices often require more support from the assistant, while experts relied on their own domain knowledge. 

\textit{Novice} queries are more \textit{similar to the exam questions}, often verbatim (despite being instructed to not copy-paste). To quantify this similarity, we measure the fuzzy string match partial ratios (Partial Sim.) between user queries and the questions they asked about. Additionally, we compute the maximum cosine similarities between sentences in the question and the query (Sem. Sim.). We assume that novice users do not know how to break down the problem to begin with due to the lack of domain knowledge. Consequently, their queries closely mirror the original exam question wording, signaling more surface-level engagement.

As a result of the similarity between non-expert queries and exam questions, novice queries now resemble the characteristics previously associated with expert queries (Section \ref{sec:expertise_classifier_desc}), opposite from our initial assumptions. Novice queries from our user study are longer, judged to exhibit higher expertise, and include more jargon. Our findings here again highlights a key limitation of passive personalization in our task context: if user expertise were inferred dynamically from query characteristics, novice queries could be misclassified as expert ones, leading to novices receiving overly concise and jargon-dense responses.

\section{Conclusion}

We investigate passive expertise-based personalization using an enterprise AI assistant for Platform A. We created a personalized assistant through prompt optimization and compared it to a standard version in a timed user study where participants completed Platform A certification exam questions with AI assistance. Our results show that personalization can moderately reduce task load, improve exam performance, and enhance perceived assistant response quality. However, in time-constrained tasks, features of expertise-oriented personalized responses like detailed examples and explanations were less helpful for non-experts, despite being preferred in less pressured settings. This suggests passive personalization using existing user expertise may not be preferable when the task is time-constrained. Therefore, it would be beneficial to \textit{combine} active and passive personalization to accommodate diverse needs and contexts for AI assistants. Overall, our work demonstrates the potential of expertise-based personalization to improve task outcomes, while highlighting the importance of user agency in interactive, task-oriented systems.

\section*{Limitations}

There are a few limitations with our work due to its exploratory nature. We aim to acknowledge and discuss these limitations here.

First, despite having a reasonable total participant pool for the user study, there are not enough participants per expertise level for us to draw stronger conclusions regarding how effective our expertise-level personalization is. One other consequence of our limited pool size is the lack of role diversity, as many individuals reported the same role (engineering). Thus, it is possible our conclusions only apply to the particular roles studied, rather than the population at large. For future work, we hope to conduct a larger-scale study with higher role diversity to better support any conclusions made.

Second, our results only pertain to the effect of passive personalization. As discussed earlier, under time constraints, user preference actually shifts towards more concise explanations despite preferring the opposite when under no time constraints. Thus, there is a need to conduct further studies on how active personalization, where users are afforded a level of control over their expertise level or other stylistic personalization dimensions, can further enhance both task performance and conversational experience.

Finally, the use of preset expertise profiles where we map roles to domain expertise does not account for potential differences in domain expertise between individuals, with the same role and level of platform experience. Because the roles are broad categories applicable to a wide range of individuals, the mapping we create does not represent the true domain expertise of everyone under a role with the same level of experience. Thus, even though two individuals may report the same role and experience level, their actual expertise in various domains may vary resulting in a better or worse experience. Again, active personalization is a potential way to address this issue. By allowing users to control their level of expertise in domains, after the mapping has been applied, responses can be further personalized to users' true domain expertise.



\bibliography{anthology,custom}

@inproceedings{qiu2019question,
  title={Question difficulty prediction for multiple choice problems in medical exams},
  author={Qiu, Zhaopeng and Wu, Xian and Fan, Wei},
  booktitle={Proceedings of the 28th acm international conference on information and knowledge management},
  pages={139--148},
  year={2019}
}

@incollection{hart1988development,
  title={Development of NASA-TLX (Task Load Index): Results of empirical and theoretical research},
  author={Hart, Sandra G and Staveland, Lowell E},
  booktitle={Advances in psychology},
  volume={52},
  pages={139--183},
  year={1988},
  publisher={Elsevier}
}

@article{lewis2020retrieval,
  title={Retrieval-augmented generation for knowledge-intensive nlp tasks},
  author={Lewis, Patrick and Perez, Ethan and Piktus, Aleksandra and Petroni, Fabio and Karpukhin, Vladimir and Goyal, Naman and K{\"u}ttler, Heinrich and Lewis, Mike and Yih, Wen-tau and Rockt{\"a}schel, Tim and others},
  journal={Advances in neural information processing systems},
  volume={33},
  pages={9459--9474},
  year={2020}
}

@article{palta2025speaking,
  title={Speaking the right language: The impact of expertise alignment in user-ai interactions},
  author={Palta, Shramay and Chandrasekaran, Nirupama and Rudinger, Rachel and Counts, Scott},
  journal={arXiv preprint arXiv:2502.18685},
  year={2025}
}

@inproceedings{opsahl-ong-etal-2024-optimizing,
    title = "Optimizing Instructions and Demonstrations for Multi-Stage Language Model Programs",
    author = "Opsahl-Ong, Krista  and
      Ryan, Michael J  and
      Purtell, Josh  and
      Broman, David  and
      Potts, Christopher  and
      Zaharia, Matei  and
      Khattab, Omar",
    editor = "Al-Onaizan, Yaser  and
      Bansal, Mohit  and
      Chen, Yun-Nung",
    booktitle = "Proceedings of the 2024 Conference on Empirical Methods in Natural Language Processing",
    month = nov,
    year = "2024",
    address = "Miami, Florida, USA",
    publisher = "Association for Computational Linguistics",
    url = "https://aclanthology.org/2024.emnlp-main.525/",
    doi = "10.18653/v1/2024.emnlp-main.525",
    pages = "9340--9366"
}

@inproceedings{ferrod2021identifying,
  title={Identifying users’ domain expertise from dialogues},
  author={Ferrod, Roger and Cena, Federica and Di Caro, Luigi and Mana, Dario and Simeoni, Rossana Grazia},
  booktitle={Adjunct Proceedings of the 29th ACM Conference on User Modeling, Adaptation and Personalization},
  pages={29--34},
  year={2021}
}

@article{toma2015tell,
  title={Tell-tale words: Linguistic cues used to infer the expertise of online medical advice},
  author={Toma, Catalina L and D’Angelo, Jonathan D},
  journal={Journal of Language and Social Psychology},
  volume={34},
  number={1},
  pages={25--45},
  year={2015},
  publisher={Sage Publications Sage CA: Los Angeles, CA}
}

@article{joshi2025eli,
  title={ELI-Why: Evaluating the Pedagogical Utility of Language Model Explanations},
  author={Joshi, Brihi and He, Keyu and Ramnath, Sahana and Sabouri, Sadra and Zhou, Kaitlyn and Chattopadhyay, Souti and Swayamdipta, Swabha and Ren, Xiang},
  journal={arXiv preprint arXiv:2506.14200},
  year={2025}
}

@article{kim2011experts,
  title={How do experts and novices differ? Relation versus attribute and thinking versus feeling in language use.},
  author={Kim, Kyungil and Bae, Jinhee and Nho, Myung-Woo and Lee, Chang Hwan},
  journal={Psychology of Aesthetics, Creativity, and the Arts},
  volume={5},
  number={4},
  pages={379},
  year={2011},
  publisher={Educational Publishing Foundation}
}

@article{flesch2007flesch,
  title={Flesch-Kincaid readability test},
  author={Flesch, Rudolf},
  journal={Retrieved October},
  volume={26},
  number={3},
  pages={2007},
  year={2007}
}

@article{dale1948formula,
  title={A formula for predicting readability: Instructions},
  author={Dale, Edgar and Chall, Jeanne S},
  journal={Educational research bulletin},
  pages={37--54},
  year={1948},
  publisher={JSTOR}
}

@article{persky2017moving,
  title={Moving from novice to expertise and its implications for instruction},
  author={Persky, Adam M and Robinson, Jennifer D},
  journal={American journal of pharmaceutical education},
  volume={81},
  number={9},
  pages={6065},
  year={2017},
  publisher={Elsevier}
}

@inproceedings{wilensky1987berkeley,
  title={The Berkeley UNIX consultant project},
  author={Wilensky, Robert},
  booktitle={Wissensbasierte Systeme: 2. Internationaler GI-Kongre{\ss} M{\"u}nchen, 20./21. Oktober 1987},
  pages={286--296},
  year={1987},
  organization={Springer}
}

@article{krippendorff2011computing,
  title={Computing Krippendorff's alpha-reliability},
  author={Krippendorff, Klaus},
  year={2011}
}

@article{rhee2020effects,
  title={Effects of personalization and social role in voice shopping: An experimental study on product recommendation by a conversational voice agent},
  author={Rhee, Chong Eun and Choi, Junho},
  journal={Computers in Human Behavior},
  volume={109},
  pages={106359},
  year={2020},
  publisher={Elsevier}
}

@inproceedings{ha2024clochat,
  title={CloChat: Understanding how people customize, interact, and experience personas in large language models},
  author={Ha, Juhye and Jeon, Hyeon and Han, Daeun and Seo, Jinwook and Oh, Changhoon},
  booktitle={Proceedings of the 2024 CHI Conference on Human Factors in Computing Systems},
  pages={1--24},
  year={2024}
}

@article{zhang2024personalization,
  title={Personalization of large language models: A survey},
  author={Zhang, Zhehao and Rossi, Ryan A and Kveton, Branislav and Shao, Yijia and Yang, Diyi and Zamani, Hamed and Dernoncourt, Franck and Barrow, Joe and Yu, Tong and Kim, Sungchul and others},
  journal={arXiv preprint arXiv:2411.00027},
  year={2024}
}

@article{ait2023power,
  title={The power of personalization: A systematic review of personality-adaptive chatbots},
  author={Ait Baha, Tarek and El Hajji, Mohamed and Es-Saady, Youssef and Fadili, Hammou},
  journal={SN Computer Science},
  volume={4},
  number={5},
  pages={661},
  year={2023},
  publisher={Springer}
}

@article{kocaballi2019personalization,
  title={The personalization of conversational agents in health care: systematic review},
  author={Kocaballi, Ahmet Baki and Berkovsky, Shlomo and Quiroz, Juan C and Laranjo, Liliana and Tong, Huong Ly and Rezazadegan, Dana and Briatore, Agustina and Coiera, Enrico},
  journal={Journal of medical Internet research},
  volume={21},
  number={11},
  pages={e15360},
  year={2019},
  publisher={JMIR Publications Toronto, Canada}
}

@article{joshi2017personalization,
  title={Personalization in goal-oriented dialog},
  author={Joshi, Chaitanya K and Mi, Fei and Faltings, Boi},
  journal={arXiv preprint arXiv:1706.07503},
  year={2017}
}

@article{li2024hello,
  title={Hello again! llm-powered personalized agent for long-term dialogue},
  author={Li, Hao and Yang, Chenghao and Zhang, An and Deng, Yang and Wang, Xiang and Chua, Tat-Seng},
  journal={arXiv preprint arXiv:2406.05925},
  year={2024}
}

@article{zheng2019personalized,
  title={Personalized dialogue generation with diversified traits},
  author={Zheng, Yinhe and Chen, Guanyi and Huang, Minlie and Liu, Song and Zhu, Xuan},
  journal={arXiv preprint arXiv:1901.09672},
  year={2019}
}

@inproceedings{joko2024doing,
  title={Doing personal laps: Llm-augmented dialogue construction for personalized multi-session conversational search},
  author={Joko, Hideaki and Chatterjee, Shubham and Ramsay, Andrew and De Vries, Arjen P and Dalton, Jeff and Hasibi, Faegheh},
  booktitle={Proceedings of the 47th International ACM SIGIR Conference on Research and Development in Information Retrieval},
  pages={796--806},
  year={2024}
}

@article{abbasian2023conversational,
  title={Conversational health agents: A personalized llm-powered agent framework},
  author={Abbasian, Mahyar and Azimi, Iman and Rahmani, Amir M and Jain, Ramesh},
  journal={arXiv preprint arXiv:2310.02374},
  year={2023}
}

@article{salemi2023lamp,
  title={Lamp: When large language models meet personalization},
  author={Salemi, Alireza and Mysore, Sheshera and Bendersky, Michael and Zamani, Hamed},
  journal={arXiv preprint arXiv:2304.11406},
  year={2023}
}

@inproceedings{li2024learning,
  title={Learning to rewrite prompts for personalized text generation},
  author={Li, Cheng and Zhang, Mingyang and Mei, Qiaozhu and Kong, Weize and Bendersky, Michael},
  booktitle={Proceedings of the ACM Web Conference 2024},
  pages={3367--3378},
  year={2024}
}

@inproceedings{luo2019learning,
  title={Learning personalized end-to-end goal-oriented dialog},
  author={Luo, Liangchen and Huang, Wenhao and Zeng, Qi and Nie, Zaiqing and Sun, Xu},
  booktitle={Proceedings of the AAAI Conference on Artificial Intelligence},
  volume={33},
  number={01},
  pages={6794--6801},
  year={2019}
}

@article{inkpen2023advancing,
  title={Advancing human-AI complementarity: The impact of user expertise and algorithmic tuning on joint decision making},
  author={Inkpen, Kori and Chappidi, Shreya and Mallari, Keri and Nushi, Besmira and Ramesh, Divya and Michelucci, Pietro and Mandava, Vani and Vep{\v{r}}ek, Libu{\v{s}}e Hannah and Quinn, Gabrielle},
  journal={ACM Transactions on Computer-Human Interaction},
  volume={30},
  number={5},
  pages={1--29},
  year={2023},
  publisher={ACM New York, NY}
}

@inproceedings{maloor2000dynamic,
  title={Dynamic user level and utility measurement for adaptive dialog in a help-desk system},
  author={Maloor, Preetam and Chai, Joyce},
  booktitle={1st SIGdial Workshop on Discourse and Dialogue},
  pages={94--101},
  year={2000}
}

@article{jokinen2004evaluation,
  title={Evaluation of adaptivity and user expertise in a speech-based e-mail system},
  author={Jokinen, Kristiina and Kanto, Kari and Kerminen, Antti and Rissanen, Jyrki},
  journal={Robust and Adaptive Information Processing for Mobile Speech Interfaces},
  pages={44},
  year={2004}
}

@article{David2025ProfiLLMAL,
  title={ProfiLLM: An LLM-Based Framework for Implicit Profiling of Chatbot Users},
  author={Shahaf David and Yair Meidan and Ido Hersko and Daniel Varnovitzky and Dudu Mimran and Yuval Elovici and Asaf Shabtai},
  journal={ArXiv},
  year={2025},
  volume={abs/2506.13980},
  url={https://api.semanticscholar.org/CorpusID:279410986}
}

@inproceedings{chen2024learning,
  title={Learning agent-based modeling with llm companions: Experiences of novices and experts using chatgpt \& netlogo chat},
  author={Chen, John and Lu, Xi and Du, Yuzhou and Rejtig, Michael and Bagley, Ruth and Horn, Mike and Wilensky, Uri},
  booktitle={Proceedings of the 2024 CHI Conference on Human Factors in Computing Systems},
  pages={1--18},
  year={2024}
}

@inproceedings{komatani2003flexible,
  title={Flexible guidance generation using user model in spoken dialogue systems},
  author={Komatani, Kazunori and Ueno, Shinichi and Kawahara, Tatsuya and Okuno, Hiroshi G},
  booktitle={Proceedings of the 41st Annual Meeting of the Association for Computational Linguistics},
  pages={256--263},
  year={2003}
}

@article{neshaei2024towards,
  title={Towards modeling learner performance with large language models},
  author={Neshaei, Seyed Parsa and Davis, Richard Lee and Hazimeh, Adam and Lazarevski, Bojan and Dillenbourg, Pierre and K{\"a}ser, Tanja},
  journal={arXiv preprint arXiv:2403.14661},
  year={2024}
}

@inproceedings{scarlatos2025exploring,
  title={Exploring knowledge tracing in tutor-student dialogues using llms},
  author={Scarlatos, Alexander and Baker, Ryan S and Lan, Andrew},
  booktitle={Proceedings of the 15th International Learning Analytics and Knowledge Conference},
  pages={249--259},
  year={2025}
}

@article{jung2024clst,
  title={CLST: Cold-Start Mitigation in Knowledge Tracing by Aligning a Generative Language Model as a Students' Knowledge Tracer},
  author={Jung, Heeseok and Yoo, Jaesang and Yoon, Yohaan and Jang, Yeonju},
  journal={arXiv preprint arXiv:2406.10296},
  year={2024}
}

@article{zhang2015predicting,
  title={Predicting users' domain knowledge in information retrieval using multiple regression analysis of search behaviors},
  author={Zhang, Xiangmin and Liu, Jingjing and Cole, Michael and Belkin, Nicholas},
  journal={Journal of the Association for Information Science and Technology},
  volume={66},
  number={5},
  pages={980--1000},
  year={2015},
  publisher={Wiley Online Library}
}

@article{kiseleva2015impact,
  title={The impact of technical domain expertise on search behavior and task outcome},
  author={Kiseleva, Julia and Garc{\'\i}a, Alejandro Montes and Kamps, Jaap and Spirin, Nikita},
  journal={arXiv preprint arXiv:1512.07051},
  year={2015}
}
\bibliographystyle{acl_natbib}

\appendix

\section{Modified Expertise Classification Prompt}
\label{app:expertise_prompt}

\begin{lstlisting}
You will be given some sentences (taken from a conversation) from a user. Your task is to determine the user's expertise based on the provided sentences.

User expertise of the topic of the conversation falls into one of the following 3 categories, which are ordered from lowest to highest level of expertise: Novice is the lowest level of expertise, followed by Intermediate, and then Expert is the highest level of expertise.

Novice: A subject novice is a person who has little or no familiarity with the topic or domain of the conversation. A subject novice may ask questions that are vague, general, irrelevant, or based on incorrect assumptions. A subject novice may also have difficulty understanding the terminology, concepts, or arguments of experts or more knowledgeable people in the subject.

Intermediate: A subject intermediate is someone who has some basic knowledge or familiarity with the topic of the conversation, but not enough to be considered an expert or even proficient. A subject intermediate can ask general questions that reflect their curiosity or interest in the topic, but not very specific or complex ones that require deeper understanding or analysis. A subject intermediate might have learned some terms or concepts related to the topic, but not how to apply them in different contexts or situations.

Expert: A subject expert is someone who can apply relevant concepts and terminology from the conversation to different scenarios and problems. They can analyze and interpret data, compare and contrast different methods or approaches, and justify their reasoning with evidence. The user also demonstrates curiosity and interest in the subject by asking questions that go beyond the surface level and explore the deeper implications and connections of the topic.

# TASK:
Q1. Classify the user expertise level based on the sentences into one of the 3 categories: Novice, Intermediate, Expert. Your answer should be a single expertise level from the provided list of expertise categories.

# ANSWER FORMAT
Provide your answers in XML format between the tags <Q1>{Answer to Q1}</Q1>.

# TIPS
- Provide all answers in **English**.
\end{lstlisting}

\section{Instruction Framework for Different Student Expertise}
\label{app:instruction_strats}

\begin{figure*}[!h]
    \centering
    \includegraphics[scale=0.3]{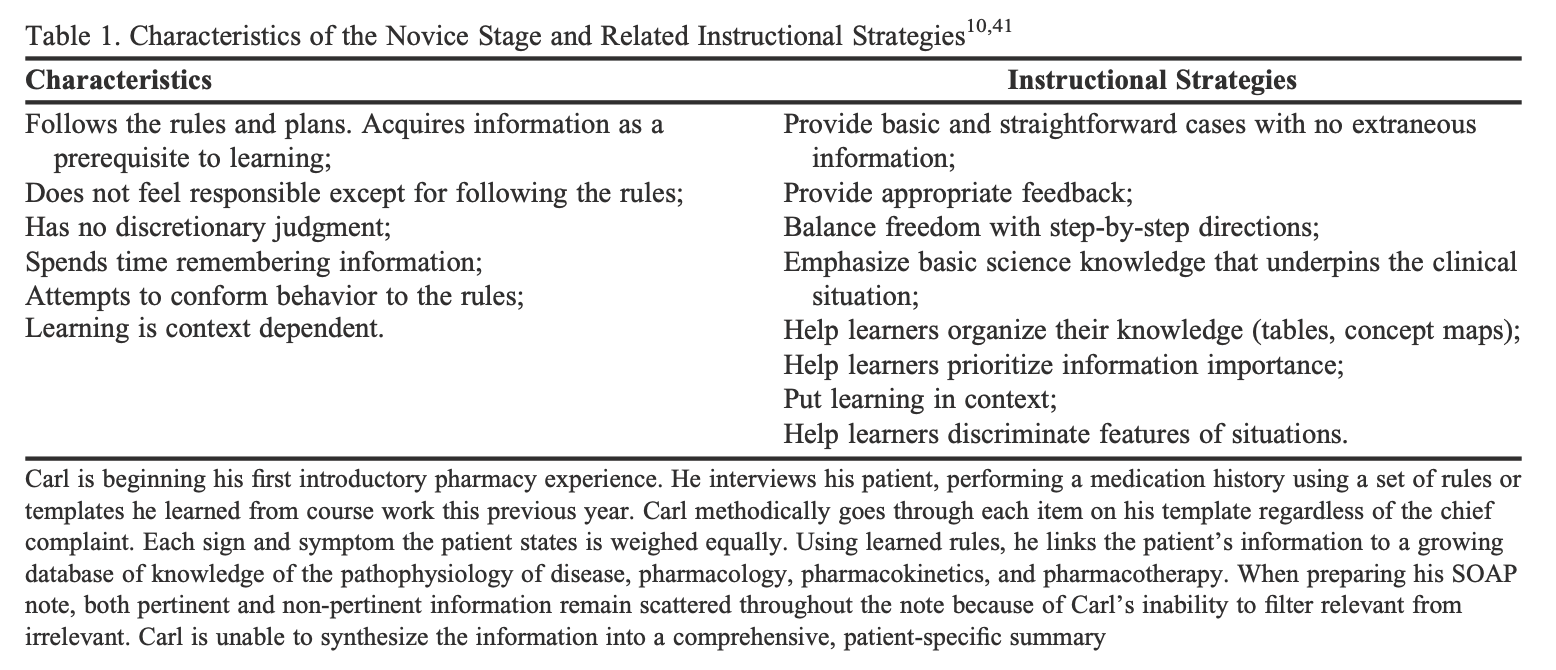}
    \caption{Instruction strategies for novice students.}
    \label{fig:novice_strat}
\end{figure*}

\begin{figure*}[!h]
    \centering
    \includegraphics[scale=0.3]{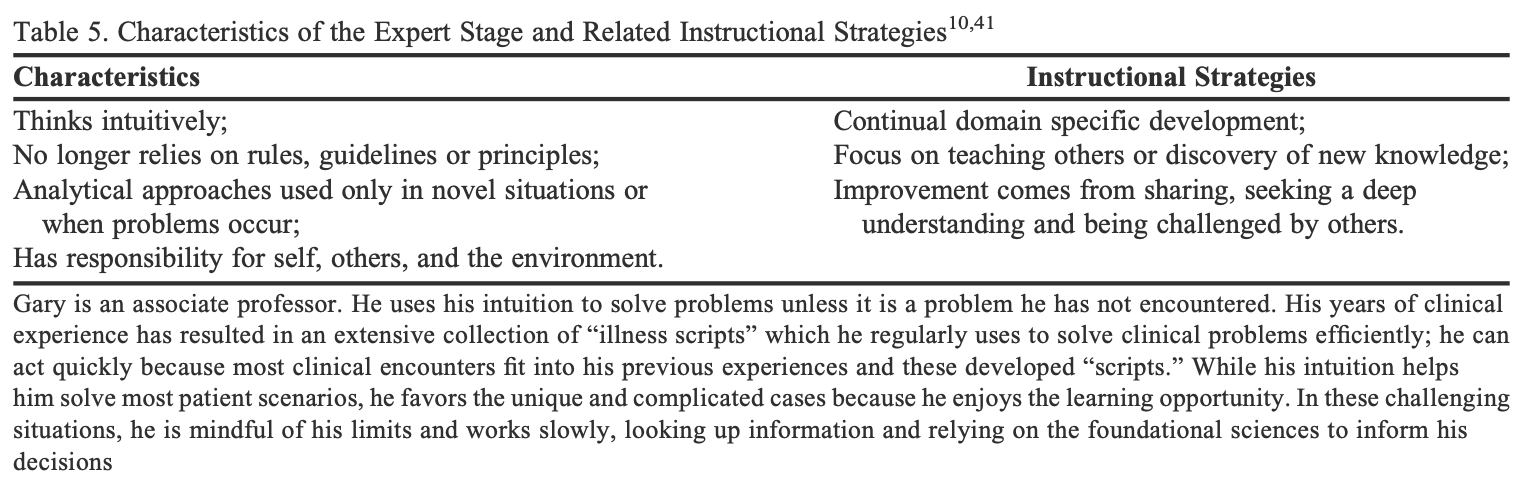}
    \caption{Instruction strategies for expert students.}
    \label{fig:exp_strat}
\end{figure*}

We copy the following instructional strategies directly from \citet{persky2017moving}. Please refer to Figures \ref{fig:novice_strat} for strategies for novice students and \ref{fig:exp_strat} for expert students.

\section{Question Preparation Process}
\label{app:material_preparation}

To select the specific 24 questions, one of the authors used the baseline assistant as the knowledge source to answer the test questions.
This author removed any questions that (1) they were not able to answer correctly with AI assistance (2) only require general common sense when answering. Essentially, this step removes overly easy questions and reduces the probability of the AI assistant being imprecise or incorrect, therefore making user performance more indicative of the effect of expertise-based personalization. 

We also needed to ensure the evenness of question difficulty among the two sets of certification questions. \citet{qiu2019question} utilized a two-dimensional model for multiple choice question difficulty, \textbf{Recall Difficulty} (how difficult is it to recall the relevant facts) and \textbf{Confusion Difficulty} (how difficult is it to differentiate between the correct answer and the distractor answers). The average of these two ratings would be the difficulty scores. We employ this framework to recruit two expert Platform A users to rate each of the questions on both dimensions on a scale of 1-5.

Given a multiple choice question and its possible choices with the correct response(s) \textbf{bolded}, expert users are asked to rate the question on a five-point Likert scale:

\begin{enumerate}
    \item Very Easy - anyone with basic Platform A knowledge can do it
    \item Easy - anyone with limited Platform A experience can do it with minimal guidance
    \item Moderate - a competent user can do it without major difficulty, though it may take some effort
    \item Challenging - requires substantial experience
    \item Very challenging - only Platform A experts can do it
\end{enumerate}

After the first round of expert ratings, the Krippendorff's $\alpha$ \cite{krippendorff2011computing} was low ($0.137$). Further inspection of our ratings indicates that the primary disagreements originate from recall difficulty (Krippendorff's $\alpha = -0.118$). We surmise this is again due to the highly diverse user population of Platform A; expert users with different professional roles differ in types of information they can recall. The two experts then resolved disagreements through discussions. We discarded questions that they still had significant disagreements on. This leaves a total of 18 questions, and the resulting Krippendorff's $\alpha$ is $0.899$.

Finally, to select questions for each condition's question set, we take the averages of the difficulty ratings from both experts, and then evenly distributed questions of similar difficulty ratings between the two sets to the best of our ability. Here are the difficulty scores for all questions in each set:
\begin{enumerate}
    \item \textbf{Question Set 1:} 1, 1.5, 1.75, 2, 2.25, 3, 3.25, 3.75
    \item \textbf{Question Set 2:} 1, 1.5, 1.75, 2, 2.25, 3.25, 3.25, 3.75
\end{enumerate}

\section{User Study Results by Expertise Level}

\subsection{Conversational Experience}

\begin{table}[!h]
\centering
\begin{tabular}{c|ccc}
\toprule
    \textbf{Cond.} & \textbf{Novice} & \textbf{Intermediate} & \textbf{Expert}\\
    \midrule
    B & \textbf{3.60} & \textbf{3.00} & 2.25\\
    P & 3.00 & 2.75 & \textbf{3.00}\\
    \bottomrule
\end{tabular}
\caption{The average expertise alignment ratings for the baseline (B) and personalized (P) assistants, across different user expertise levels.}
\label{tab:assistant_response_exp}
\end{table}

\subsection{Task Performance}
\label{app:task_performance_ne}

\begin{figure*}[!ht]
    \centering
    \includegraphics[scale=0.5]{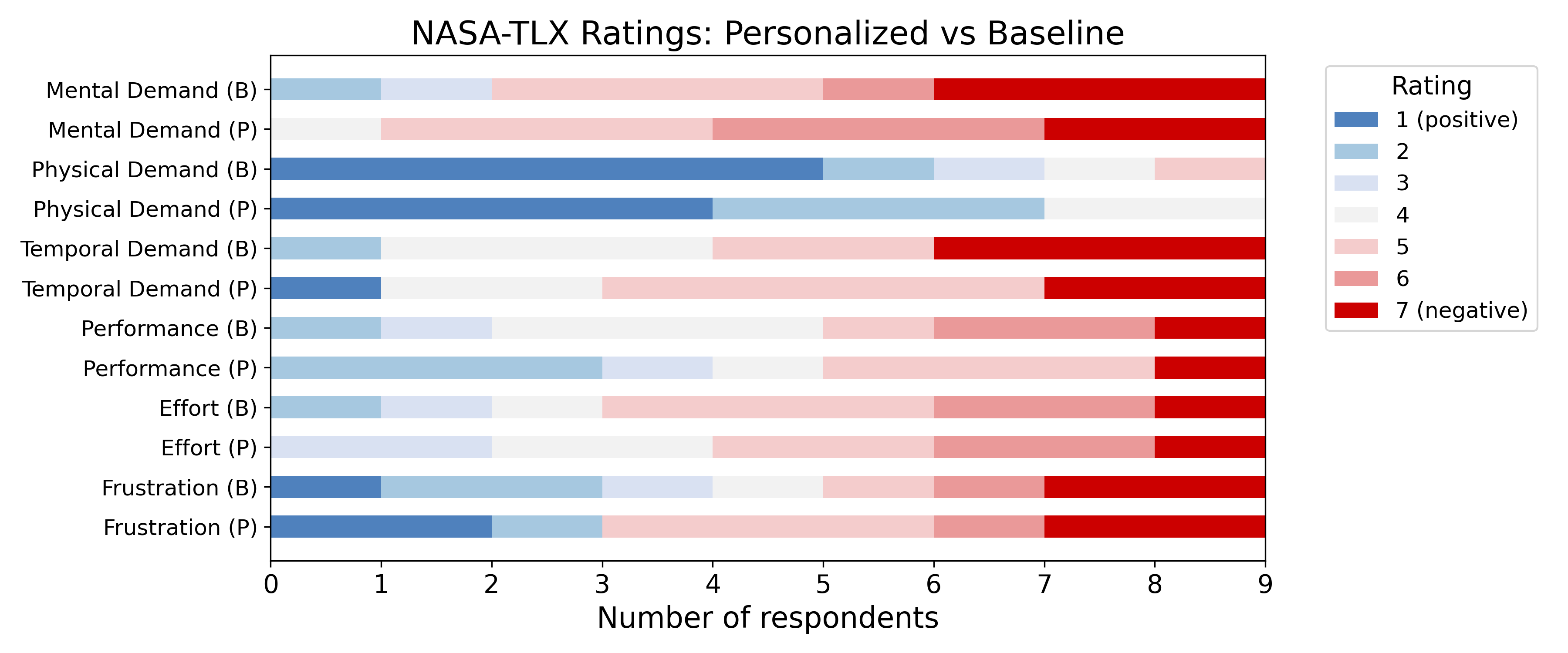}
    \begin{tabular}{c|cccccc}
    \toprule
         \textbf{Condition} & \textbf{Mental} $\downarrow$ & \textbf{Physical} $\downarrow$ & \textbf{Temporal} $\downarrow$ & \textbf{Performance} $\downarrow$& \textbf{Effort} $\downarrow$& \textbf{Frustration} $\downarrow$\\
         \midrule
         Baseline & \textbf{5.22} & 2.11 & 5.00 & 4.56 & 4.78 & \textbf{4.11}\\
         Personalized & 5.66 & \textbf{2.00} & \textbf{4.78} & \textbf{3.89}& 4.78& 4.33\\
         \bottomrule
    \end{tabular}
    \caption{NASA-TLX results for baseline (B) and our personalized system (P) for the nine non-experts participants. We report the average ratings in the table.}
    \label{fig:nasa_tlx_ne}
\end{figure*}

We record the NASA-TLX results for non-experts in Figure \ref{fig:nasa_tlx_ne}. We notice similar trends compared to the results when including all participants.

\end{document}